\documentclass[aps,prl,reprint,superscriptaddress,nofootinbib,floatfix]{revtex4-2}

\usepackage{amsmath,amssymb,mathtools}
\usepackage{bm}
\usepackage{tikz}
\usetikzlibrary{arrows.meta,positioning,fit,calc}
\usetikzlibrary{arrows.meta,positioning}
\usepackage[hidelinks]{hyperref}
\begin{document}

\title{Latent Geometry and the Emergence of Lorentzian Time in Matrix Quantum Mechanics}

\author{Badis Ydri}
\affiliation{Department of Physics, Badji Mokhtar Annaba University, Algeria}

\begin{abstract}

We introduce \emph{latent geometry} in BFSS/BMN models.  Before an emergent geometry evaporates at the geometric transition,
large-$d$ Gaussianization encodes its information in the exceptional
$N=2$ sector selected by the baby fuzzy sphere, a single active spin-$1/2$
Pauli block.   Dimensional reduction,
finite-$N$ Gaussianization and exact Molien--Weyl projection transport
this encoding to mass-deformed BFSS$_2$.  At $N=2$, the same stable BFSS$_2$ oscillator admits inequivalent
unitary $\mathbf{H}^2_\theta=\mathbf{EAdS}^2_\theta$ and
$\mathbf{dS}^2_\theta$ decodings, realizing spacetime transmutation
and the emergence of Lorentzian time.

\end{abstract}

\maketitle

\medskip
\noindent
Matrix models place quantum dynamics and noncommutative geometry in the same microscopic variables.  The BFSS model and its BMN deformation are central examples~\cite{BFSS,BMN}; the BMN/Myers interaction supports fuzzy-sphere configurations~\cite{Myers}.  The fuzzy sphere itself goes back to Hoppe and Madore~\cite{Hoppe,Madore}, and Yang--Mills--Chern--Simons matrix models exhibit its nonperturbative phase dynamics and the transition between matrix and geometric phases~\cite{AzumaFuzzy,GeometryTransitionPRL,MatrixModelsEmergent,SteinackerEmergentGeometry}.  We ask what becomes of the geometric information when this macroscopic order is absent in the Yang--Mills phase.  Must geometry be regarded as lost, or can it remain encoded in the matrix system?

\medskip
\noindent
The question of Lorentzian spacetime has a separate history in matrix
models.  In the Lorentzian type-IIB/IKKT formulation, real-time
dynamics and an expanding three-dimensional space arise directly from
a Lorentzian matrix model~\cite{KimNishimuraTsuchiya}.  The mechanism
considered here is different.  For the representative construction we
begin with a Euclidean noncommutative geometry, allow its macroscopic
realization to disappear, and ask whether the surviving quantum matrix
data can subsequently support a Lorentzian geometric realization.
The resulting notion of emergent time therefore does not require a
Lorentzian microscopic action from the outset.

\medskip
\noindent
We realize this second possibility through a four-step
geometry--matrix--geometry circuit.  Step~I is the familiar dynamical
condensation, or emergence, of macroscopic geometry in the geometric
phase, where the Yang--Mills--Myers dynamics supports the fuzzy sphere.
Step~II identifies, within the same geometric phase, the baby fuzzy
sphere---the single spin-$1/2$ Pauli block that survives large-$d$
Gaussianization---as the minimal \emph{geometric encoding} of the
original fuzzy geometry.  Its existence provides the geometric
justification for restricting the same matrix model to the exceptional
$N=2$ sector.  Step~III transports this encoding through the
exceptional $N=2$ matrix-quantum-mechanical dynamics associated with
the Yang--Mills phase.  Step~IV instead realizes geometric decoding:
the $\mathfrak{so}(1,2)\simeq\mathfrak{su}(1,1)$ structure of the
mass-deformed BFSS$_2$ Hilbert space quantizes a two-dimensional
coadjoint-orbit phase space, yielding a noncommutative
constant-curvature geometry.  At the exceptional $N=2$ endpoint, the
same stable oscillator admits the two inequivalent unitary realizations
$\mathbf{H}^2_\theta=\mathbf{EAdS}^2_\theta$ and $\mathbf{dS}^2_\theta$. Steps~I and IV realize two complementary mechanisms of quantized geometry
in the present circuit.  Step~I uses a matrix-model \emph{second
quantization of geometry}---phase dynamics, or dynamical/emergent
noncommutative geometry---and is traversed in reverse as geometric
evaporation, whereas Step~IV uses a \emph{first quantization of
geometry}---geometric quantization, or static noncommutative
geometry---and is traversed forward as geometric decoding. The technical development of Steps~II--III and Step~IV is given in companion works~\cite{YdriLatentTransport,YdriBFSS2Latent};
here we isolate the minimal argument and its two linked consequences:
\emph{latent geometry} and the \emph{emergence of Lorentzian time}.
The logical structure is

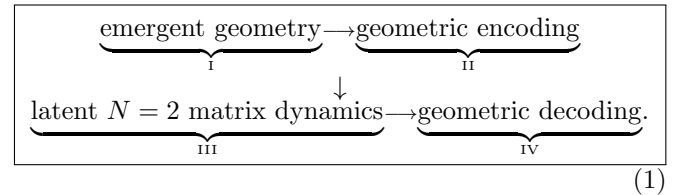
\begin{equation}
\boxed{
\begin{array}{c}
{\scriptstyle \underbrace{{\rm emergent\ geometry}}_{\rm I}
\longrightarrow
\underbrace{{\rm geometric\ encoding}}_{\rm II}}
\\[1mm]
\downarrow
\\[-1mm]
{\scriptstyle \underbrace{{\rm latent}\ N=2\ {\rm matrix\ dynamics}}_{\rm III}
\longrightarrow
\underbrace{{\rm geometric\ decoding}}_{\rm IV}}.
\end{array}}
\label{eq:four_steps}
\end{equation}
The output need not reproduce the input: curvature, compactness and even signature can change while the explicitly tracked algebraic information remains available.  We reserve \emph{geometric encoding} for Step~II and \emph{latent geometry} for the complete geometry--encoding--matrix-dynamics--geometry circuit.  The claim is not that a $2\times2$ block stores every metric datum of the fuzzy sphere, but that its minimal noncommutative algebraic core remains available for subsequent quantum-geometric realization.


\begin{figure}[t]
\centering
\begin{tikzpicture}[
    >=Latex,
    font=\scriptsize,
    every node/.style={align=center},
    box/.style={
        draw,
        rounded corners,
        minimum width=3.45cm,
        minimum height=0.82cm,
        inner sep=3pt
    },
    smallbox/.style={
        draw,
        rounded corners,
        minimum width=2.6cm,
        minimum height=0.72cm,
        inner sep=3pt
    },
    phasebox/.style={
        draw,
        rounded corners,
        dashed,
        inner sep=6pt
    }
]

\node[box] (ymmyers) at (0,0)
{Yang--Mills--Myers\\matrix model};

\node[box, below=7mm of ymmyers] (s2in)
{Step~I:\\fuzzy sphere $\mathbf{S}^{2}_{\theta}$ (emegrent geometry)};

\node[box, below=7mm of s2in] (pauli)
{Step~II:\\baby fuzzy sphere ($\mathrm{Mat}_{2}$ Pauli encoding};

\node[phasebox, fit=(ymmyers)(s2in)(pauli),
      label={[yshift=1mm]\scriptsize geometric phase}] (geo) {};

\node[box, below=17mm of pauli] (stepIII)
{Step~III:\\ Latent $N=2$ matrix dynamics};

\node[box, below=7mm of stepIII] (bfss2)
{Step~IV:\\ mass-deformed BFSS$_2$ singlet endpoint\\
$k=\frac34,\quad C=\frac{3}{16},\quad \Lambda=-s^2<0$};

\node[
    phasebox,
    fit=(stepIII)(bfss2),
    inner xsep=15mm,
    inner ysep=4mm,
    label={[yshift=1mm]\scriptsize Yang--Mills phase}
] (ymphase) {};


\coordinate (leftout)  at ($(bfss2.south)+(-2.40cm,0)$);
\coordinate (rightout) at ($(bfss2.south)+( 2.40cm,0)$);

\node[smallbox, anchor=north] (h2)
at ($(leftout)+(0,-6mm)$)
{$\mathbf{H}^{2}_{\theta}$ (discrete $D^+_{3/4}$)\\
spacetime transmutation};

\node[smallbox, anchor=north] (ds2)
at ($(rightout)+(0,-6mm)$)
{$\mathbf{dS}^{2}_{\theta}$ (complementary)\\
emergence of Lorentzian time};

\node[smallbox, below=6mm of ds2] (s2out)
     {$\mathbf{S}^{2}_{\theta}$\\
       geometric recovery};

\node[smallbox, below=6mm of h2] (ads2)
{$\mathbf{AdS}^{2}_{\theta}$\\};

\draw[->,line width=2pt] ([xshift=-2.0cm]geo.south) --
node[midway,left]{\scriptsize geometric transition}
([xshift=-2.0cm]ymphase.north);

\draw[->] (leftout)  -- (h2.north);
\draw[->] (rightout) -- (ds2.north);

     \draw[<->,dashed]
(ds2.south) -- node[midway,left]{Wick continuation} (s2out.north);

\draw[<->,dashed]
(h2.south) -- node[midway,left]{Wick continuation} (ads2.north);

\end{tikzpicture}
\caption{Latent-geometry circuit: geometric evaporation, geometric encoding, $N=2$ transport,
and BFSS$_2$-induced geometric decoding, with Lorentzian-time emergence,
recovery and spacetime transmutation.}
\label{fig:latent_circuit}
\end{figure}
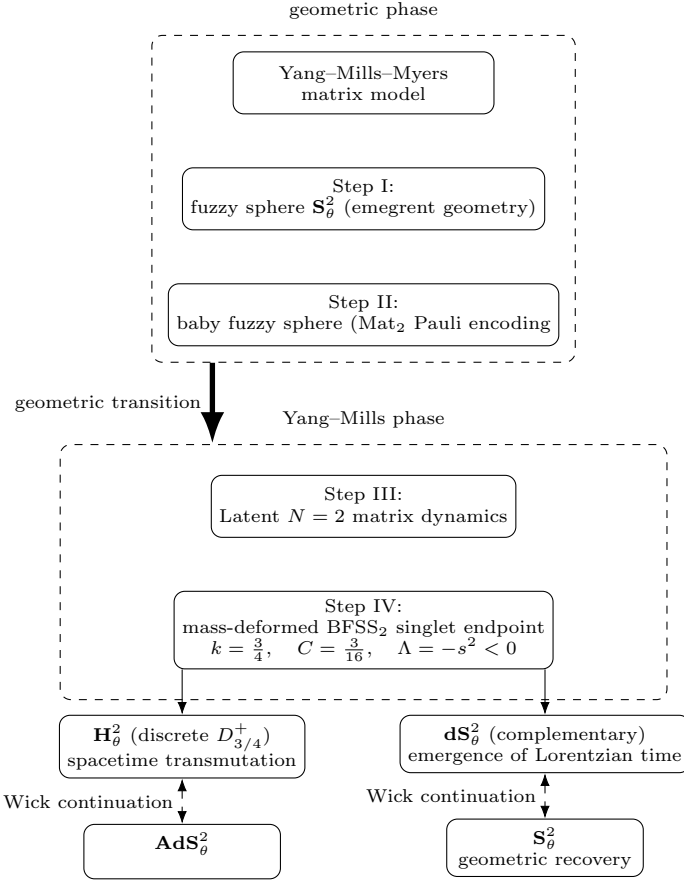

\medskip
\noindent
\emph{Encoding the fuzzy sphere.—}
For the three geometrically active directions, consider the bosonic Yang--Mills--Myers sector
\begin{equation}
\begin{aligned}
S=N\!\int_0^\beta\!dt\,{\rm Tr}\Big[&
\frac12(D_tX_A)^2-\frac14[X_A,X_B]^2\\
&+\frac{i\kappa}{3}\epsilon_{abc}X_aX_bX_c\Big].
\end{aligned}
\label{eq:YMMyers}
\end{equation}
with $A=1,\ldots,d$ and $a,b,c=1,2,3$.  It supports the irreducible fuzzy sphere $X_a=rJ_a$, $[J_a,J_b]=i\epsilon_{abc}J_c$, the standard finite-dimensional noncommutative sphere of Hoppe and Madore, whose nonperturbative matrix-model phase dynamics was studied already in Ref.~\cite{AzumaFuzzy}; see also Refs.~\cite{Hoppe,Madore,Myers,GeometryTransitionPRL,MatrixModelsEmergent}.

\medskip
\noindent
At large $d$, the Yang--Mills interaction admits a self-consistent
Gaussian description
\cite{MandalMahatoMorita,FilevOConnorBFSS,YdriLargeD,YdriLargeDSaddle}.
To leading order the commutator interaction is replaced by a dynamically
generated quadratic potential,
\begin{equation}
-\frac14{\rm Tr}[X_A,X_B]^2
\longrightarrow
\frac{\Omega^2}{2}{\rm Tr}X_A^2,
\qquad
\Omega^3\simeq d,
\label{eq:large_d_gaussian}
\end{equation}
in the massless normalization used here.  More generally, the
mass-deformed large-$d$ saddle obeys
$\Omega^3-m\Omega=d$~\cite{YdriLargeDSaddle}.  Thus the detailed angular
commutator stiffness of the Yang--Mills interaction is replaced by
radial matrix-harmonic-oscillator dynamics.

\medskip
\noindent
In the geometric sector, this Gaussianization was found in
Ref.~\cite{YdriLargeD} not to preserve the macroscopic irreducible fuzzy
sphere.  Instead, only a three-cut remnant survives whose nontrivial
component is a single spin-$1/2$ Pauli block, introduced there as the
\emph{baby fuzzy sphere}.  This configuration is in fact selected by a
sharp global variational bound, without assuming a fuzzy-sphere ansatz.
Define
$A={\rm Tr}\sum_aX_a^2$ and
$C=-i\epsilon_{abc}{\rm Tr}X_aX_bX_c$.  Cyclicity,
Hilbert--Schmidt Cauchy--Schwarz, and the sharp
B\"ottcher--Wenzel inequality~\cite{BottcherWenzel} give
\begin{equation}
\begin{split}
|C|
&=
3\left|{\rm Tr}\,X_1[X_2,X_3]\right|
\\
&\leq
3\,\|X_1\|_F\,\|[X_2,X_3]\|_F
\leq
3\sqrt2\,x_1x_2x_3
\leq
\sqrt{\frac23}\,A^{3/2},
\end{split}
\label{eq:sharp_bound}
\end{equation}
where $x_a^2=\|X_a\|_F^2={\rm Tr}X_a^2$ and the last step follows
from $x_1x_2x_3\leq(A/3)^{3/2}$.

\medskip
\noindent
The bound is saturated by a single embedded Pauli block,
\begin{equation}
\boxed{
X_a
=
\sqrt{\frac{2A}{3}}\,
\frac{\sigma_a}{2}\oplus0_{N-2},
\qquad
\frac{N-1}{2}
\longrightarrow
\frac12\oplus0^{\oplus(N-2)},
}
\label{eq:pauli_encoder}
\end{equation}
for which, with $\mu=\sqrt{2A/3}$,
\begin{equation}
A=\frac32\mu^2,
\qquad
C=\frac32\mu^3,
\qquad
\frac{C}{A^{3/2}}=\sqrt{\frac23}.
\label{eq:pauli_saturation}
\end{equation}
Thus the Pauli block saturates every inequality entering
Eq.~\eqref{eq:sharp_bound} and is a global maximizer of $C$ at fixed
quadratic radius.  Moreover, $q$ identical Pauli blocks give
$C_q(A)=\sqrt{2/3}\,A^{3/2}/\sqrt q$, so the cubic gain is maximal for
$q=1$.  \medskip
\noindent
Consequently, for any radially stabilized cubic potential
$V[X]=V_{\rm rad}(A)-\gamma C$ with $\gamma>0$, the exact angular
minimization gives
\begin{equation}
V_{\rm eff}(A)
=
V_{\rm rad}(A)
-\gamma\sqrt{\frac23}\,A^{3/2}.
\label{eq:pauli_radial_reduction}
\end{equation}
The angular minimization is therefore solved globally by the single
Pauli block. The macroscopic irreducible sphere is therefore replaced by its minimal
noncommutative $SU(2)$ carrier.  In the present construction we interpret
this baby fuzzy sphere as the minimal \emph{geometric encoding} of the
original fuzzy geometry~\cite{YdriLatentTransport}.  Its minimal
standalone realization is $N=2$, providing the first $N=2$
exceptionality before the subsequent matrix quantum mechanics is
invoked.

\medskip
\noindent
This identification also determines the degrees of freedom that must
be retained in the latent dynamics.  The encoded $SU(2)$ geometry
occupies only the three Myers directions, whereas the remaining
$d-3$ matrices are geometrically inactive but continue to determine
the self-consistent quadratic scale.  The relevant latent sector is
therefore the $N=2$ three-matrix BFSS$_4$/BMN$_4$ core, with the
large-$d$ background retained through its induced mass.

\medskip
\noindent
\emph{Transport through the $N=2$ matrix dynamics.—}
In the Yang--Mills phase, where the Myers term has vanished, the encoded
$N=2$ three-matrix sector admits three successive operations.  First, one active matrix is exactly Gaussian for arbitrary configurations of the other two and can be integrated out.  If $X_3$ is eliminated, its complete contribution is
\begin{equation}
\Gamma_3[X_1,X_2,A_t]
=
\frac12{\rm Tr}_{t,{\rm adj}}
\log\!\left[-D_t^2+m+{\rm ad}_{X_1}^2+{\rm ad}_{X_2}^2\right].
\label{eq:exact_X3_core}
\end{equation}
Equation~\eqref{eq:exact_X3_core} is exact in $X_1$ and $X_2$.  At low temperature its nonzero-winding sectors are exponentially suppressed, while the zero-winding part admits a local expansion around the MHO saddle inherited from the large-$d$ theory~\cite{YdriLatentTransport}.

\medskip
\noindent
Second, the resulting two-matrix BFSS$_3$ theory is organized around a finite-$N$ Gaussian saddle, following the mechanism established by Pavel for $SU(2)$ Yang--Mills quantum mechanics~\cite{PavelYM}.  Writing $\Omega_3>0$ for the inherited BFSS$_3$ MHO gap, determined by
$\Omega_3^3-m\Omega_3=3$, and $g^2$ for the quartic coupling, the
optimized frequency obeys

\begin{equation}
\omega_2^3-\Omega_3^2\omega_2-g^2=0,
\qquad
\epsilon_{\rm G}=\frac{g^2}{\omega_2^3}
\simeq\frac{g^2}{\Omega_3^3}.
\label{eq:gaussian_control_core}
\end{equation}
Thus the Gaussianization becomes increasingly accurate as the inherited
mass grows, while remaining already effective in the massless case.
This is verified independently by the Rayleigh--Ritz analysis in Appendix ~\ref{app:numerics}.

\medskip
\noindent Third, once the BFSS$_3$ MHO is reached, the Gauss law is imposed exactly by Molien--Weyl projection:
\begin{equation}
\boxed{
\begin{array}{c}
(2,3)\,{\rm BFSS}_4
\longrightarrow
(2,2)\,{\rm BFSS}_3^{\rm eff}
\\[1mm]
\longrightarrow
(2,2)\,{\rm BFSS}_3{\rm -MHO}
\longrightarrow
\left({\cal H}^{\rm sing}_{\rm BFSS_2}\right)^{\otimes3}.
\end{array}}
\label{eq:matrix_chain}
\end{equation}
Appendix~\ref{app:stepIII} derives the exact endpoint.  For two $SU(2)$ adjoint oscillators the invariant ring is freely generated by the three quadratic Gram invariants
$G_{11}=\mathrm{Tr}\,X_1^2$, $G_{22}=\mathrm{Tr}\,X_2^2$, and
$G_{12}=\mathrm{Tr}(X_1X_2)$; no independent cubic invariant survives, and higher invariants reduce to Gram products. Hence~\cite{OConnor2023,OConnor2024,YdriMolienWeyl}
\begin{equation}
\boxed{
Z^{\rm bos}_{2,2}(x)=\frac{1}{(1-x^2)^3}
=\left[Z^{\rm bos}_{2,1}(x)\right]^3.
}
\label{eq:MWfactor}
\end{equation}
The tensor cube denotes three singlet towers in one commonly gauged $SU(2)$ system, not a Cartesian product of spacetimes.  Each tower carries the same BFSS$_2$ representation mechanism.  This is a second, independent $N=2$ exceptionality: the Pauli encoder is minimal at $N=2$, while two adjoint $SU(2)$ oscillator vectors possess exactly the three free Gram generators required for the cube.

\medskip
\noindent
\emph{BFSS$_2$ as a geometric endpoint.—}
Each factor reached in Eq.~\eqref{eq:matrix_chain} is the gauged mass-deformed BFSS$_2$ matrix harmonic oscillator,
\begin{equation}
H=\frac12{\rm Tr}(P^2+s^2X^2),
\qquad
\Lambda=-s^2<0.
\label{eq:BFSS2}
\end{equation}
Physical states obey the Gauss law $G=i[P,X]=0$.  The quadratic singlets generate $\mathfrak{so}(1,2)\simeq\mathfrak{su}(1,1)$~\cite{Park2005,Kim2006}.  With $M=N^2-1$ adjoint components, the Fock vacuum is a physical lowest-weight state with
\begin{equation}
k_0=\frac{N^2-1}{4},
\qquad
C_{\mathfrak{so}(1,2)}=-k_0(k_0-1).
\label{eq:general_k}
\end{equation}
For $N\geq3$, $k_0\geq2$ and the vacuum Casimir is strictly negative, giving the ordinary positive-energy discrete-series branch.  At $N=2$, however,
\begin{equation}
\boxed{
k=\frac34,
\qquad
C_{\mathfrak{so}(1,2)}=\frac{3}{16},
\qquad
\Lambda=-s^2<0.}
\label{eq:N2data}
\end{equation}
At these exceptional values, the same stable $N=2$ BFSS$_2$
oscillator admits two inequivalent unitary global realizations of its
$\mathfrak{so}(1,2)$ algebra.  The lowest-weight discrete series
$D^+_{3/4}$ gives
$\mathbf{H}^2_\theta=\mathbf{EAdS}^2_\theta$, while the complementary
series at the same Casimir value $C=3/16$ gives the inequivalent
$\mathbf{dS}^2_\theta$ realization
~\cite{Bargmann1947,HoLiLargeN,JurmanSteinacker,PinzulStern,
YdriBFSS2Latent,Ydri:2021AdS2I,Ydri:2021AdS2II,Bouraiou:2021AdS2III}. The endpoint data themselves do not select one of these unitary realizations over the other.  Hence
\begin{equation}
  \boxed{
{\rm BFSS}_2
\Big|_{N=2,\Lambda=-s^2}
\Longrightarrow
\left\{\mathbf{H}^2_\theta,\mathbf{dS}^2_\theta\right\}.
}
\label{eq:decoder}
\end{equation}
Appendix~\ref{app:decoder} gives the derivation.   The two outputs are inequivalent
unitary representations: $D^+_{3/4}$ has a lowest-weight state,
whereas the complementary series does not, and their global
realization and $\ast$--structure differ.  Both are nevertheless
admitted by the same stable $N=2$ oscillator with fixed
$\Lambda=-s^2<0$; the de~Sitter channel is therefore not an
inverted-oscillator branch.  No dynamical criterion privileges one
realization over the other.

\medskip
\noindent
The double unitary realization is exceptional to $N=2$.  At this
matrix size,
$k=3/4$ and $C=3/16$ lie simultaneously in the lowest-weight
discrete-series and complementary-series unitary domains.  For
$N\geq3$, $k_0\geq2$ and the complementary-series possibility
disappears.  Thus the same matrix size that realizes the minimal
Pauli encoding is also the unique non-Abelian endpoint admitting both
hyperbolic and de~Sitter geometric realizations.

\medskip
\noindent
\emph{Why the $N=2$ circuit closes.—}
Three independent facts coincide at the smallest non-Abelian matrix
size: Step~II selects the spin-$1/2$ Pauli carrier, Step~III yields the
three-generator $SU(2)$ singlet ring and the BFSS$_2$ towers, and
Step~IV fixes the exceptional weight $k=3/4$.  Their conjunction turns
the baby fuzzy sphere from a Gaussian remnant into a geometric encoder
whose $N=2$ quantum dynamics closes on a noncommutative geometric
sector.

\medskip
\noindent
The two geometric endpoints nevertheless arise through different notions
of quantization in the present circuit.  Step~I realizes a matrix-model
\emph{second quantization of geometry}: the quantum matrix dynamics
possesses distinct Yang--Mills and geometric phases, and the fuzzy sphere
condenses or evaporates across the geometric transition.  Step~IV instead
realizes a \emph{first quantization of geometry}: the
$\mathfrak{so}(1,2)$ structure of the mass-deformed BFSS$_2$ Hilbert
space quantizes the corresponding two-dimensional coadjoint-orbit phase
space, whose coordinate functions become noncommuting operators.  The
resulting $\mathbf{H}^2_\theta=\mathbf{EAdS}^2_\theta$ and
$\mathbf{dS}^2_\theta$ are therefore noncommutative quantized geometries,
rather than dynamical phases of BFSS$_2$ in this construction.

\medskip
\noindent
The two steps are also traversed oppositely by the information flow of
the latent-geometry circuit.  The dynamical direction of Step~I is from
matrices to emergent geometry, whereas the circuit traverses it in
reverse, encoding the emergent geometry into its latent $N=2$ matrix
carrier.  Step~IV is traversed forward as geometric decoding: the latent
BFSS$_2$ quantum data are mapped to either the discrete-series
hyperbolic realization or the complementary-series de~Sitter
realization.

\medskip
\noindent
\emph{Latent geometry and Lorentzian time.—}
Combining Eqs.~\eqref{eq:pauli_encoder}, \eqref{eq:matrix_chain} and
\eqref{eq:decoder} gives the two primary latent transitions
\begin{equation}
\boxed{
\mathbf{S}^2_\theta\leadsto\mathbf{H}^2_\theta,
\qquad
\mathbf{S}^2_\theta\leadsto\mathbf{dS}^2_\theta.
}
\label{eq:primary}
\end{equation}
Here $\leadsto$ denotes the complete transition through geometric
evaporation, encoding, $N=2$ matrix dynamics and geometric decoding
within the Yang--Mills phase, and is therefore not an ordinary Wick
rotation.  The first branch changes a compact positively curved
Euclidean geometry into a noncompact negatively curved one; the second
ends in a Lorentzian de~Sitter realization.  This differs from the direct
continuation
\begin{equation}
\mathbf{S}^2_\theta\longleftrightarrow\mathbf{dS}^2_\theta,
\label{eq:direct_wick_core}
\end{equation}
between already manifest geometric real forms.  In the latent channel
the sphere is compressed to the Pauli algebra and transported through
the $N=2$ quantum mechanics before the same stable BFSS$_2$ oscillator
admits a Lorentzian geometric realization.  This intermediate
matrix-dynamical stage gives the emergence of Lorentzian time its
dynamical content.

\medskip
\noindent
The microscopic time parameter of matrix quantum mechanics is not
created, and neither unitary realization is dynamically privileged.
Both are admitted by the same stable $N=2$ BFSS$_2$ oscillator; the
complementary-series realization makes manifest a Lorentzian
\emph{geometric} time direction.

\medskip
\noindent
The same circuit also permits \emph{geometric recovery}: the encoded
geometry may reappear as a noncommutative geometric realization within
the Yang--Mills phase, without restoring the original fuzzy-sphere
ordered phase.  Standard continuations relate
\begin{equation}
\mathbf{dS}^2_\theta\longleftrightarrow\mathbf{S}^2_\theta,
\qquad
\mathbf{H}^2_\theta\longleftrightarrow\mathbf{AdS}^2_\theta.
\label{eq:continuations}
\end{equation}
These continuations extend the latent channels to the family
$\{\mathbf{S}^2_\theta,\mathbf{H}^2_\theta,
\mathbf{dS}^2_\theta,\mathbf{AdS}^2_\theta\}$. The continuation web is
\begin{equation}
\begin{array}{ccc}
\mathbf{S}^2_\theta & \longleftrightarrow & \mathbf{dS}^2_\theta
\\[1mm]
\updownarrow && \updownarrow
\\[1mm]
\mathbf{H}^2_\theta=\mathbf{EAdS}^2_\theta
& \longleftrightarrow & \mathbf{AdS}^2_\theta.
\end{array}
\label{eq:geometry_web_core}
\end{equation}
The horizontal arrows are Euclidean--Lorentzian continuations and
the vertical relations connect the corresponding compact/noncompact or
space--time-exchanged real forms.  These continuations enlarge the
geometric output but are not the dynamical mechanism of
Eq.~\eqref{eq:primary}.  In particular, we have
\begin{equation}
\boxed{
\mathbf{S}^2_\theta
\leadsto
\mathbf{dS}^2_\theta
\longleftrightarrow
\mathbf{S}^2_\theta.
}
\label{eq:recovery}
\end{equation}
This recovers the entrance geometry as a decoded noncommutative geometry in
the Yang--Mills phase, rather than as the original fuzzy-sphere
condensate.  More generally, define
\begin{equation}
{\mathfrak G}_\theta=
\{\mathbf{S}^2_\theta,\mathbf{H}^2_\theta,
\mathbf{dS}^2_\theta,\mathbf{AdS}^2_\theta\}.
\label{eq:family}
\end{equation}
Any ${\cal G}_{\rm in}\in{\mathfrak G}_\theta$ may first be related by
standard continuation to the fuzzy-sphere representative, after which
the latent transition acts,
\begin{equation}
{\cal G}_{\rm in}
\xleftrightarrow{\rm continuation}
\mathbf{S}^2_\theta\big|_{\rm geom}
\leadsto
{\cal G}_{\rm out}\big|_{\rm YM},
\qquad
{\cal G}_{\rm out}\in{\mathfrak G}_\theta.
\label{eq:general_latent_map}
\end{equation}
Thus the construction maps the constant-curvature family into itself.

\medskip
\noindent
Recovery is the closed channel in Eq.~\eqref{eq:recovery}; generically
the output is transmuted.  The $\mathbf{H}^2_\theta$ branch changes
curvature and compactness, while the $\mathbf{dS}^2_\theta$ branch
changes signature and opens a Lorentzian time direction; space--time
exchange also reaches $\mathbf{AdS}^2_\theta$.  Thus ``No geometric
loss'' does not mean conservation of curvature, signature, compactness
or topology, but that the tracked algebraic geometric information is
encoded rather than discarded and remains available for later
geometric realization.

\medskip
\noindent
The exact and effective ingredients remain distinct.  Pauli selection
is exact within the radially stabilized cubic problem, while the
large-$d$ Gaussianization leading to it is effective.  In Step~III,
$X_3$ integration is exact before the low-temperature expansion,
finite-$N$ Gaussianization is controlled but approximate, and
Molien--Weyl factorization is exact at the MHO endpoint.  In Step~IV,
the discrete- and complementary-series realizations remain on equal
representation-theoretic footing.

\medskip
\noindent
A macroscopic geometry can therefore evaporate in the Yang--Mills phase
while its algebraic content survives latently, allowing recovery or
transmutation.  In particular, the de~Sitter branch provides a
matrix-quantum-mechanical route from Euclidean geometry to emergent
Lorentzian time.  Open questions include dynamical selection between
the two unitary realizations, extension beyond $N=2$, and survival of
the circuit in the full interacting BFSS/BMN theory.

\medskip
\noindent
\emph{Acknowledgments.—}
The author acknowledges the use of ChatGPT-5.6 for
language editing, LaTeX generation, symbolic checks, reference searches,
and as an artificial sounding board for organizing and refining ideas.
The scientific vision, design, final editing, and all intellectual
responsibility remain solely with the author.

\appendix
\setcounter{secnumdepth}{1}

\section{Reduction and exact singlet factorization}
\label{app:stepIII}

\medskip
\noindent
The first reduction in Step~III is exact before any local expansion.  For fixed $X_1$, $X_2$ and $A_t$, the eliminated matrix $X_3$ appears quadratically,
\begin{equation}
\begin{aligned}
S_3&=\int_0^\beta dt\,{\rm Tr}\left(X_3{\cal K}_3X_3\right),\\
{\cal K}_3&=-D_t^2+m+{\rm ad}_{X_1}^2+{\rm ad}_{X_2}^2.
\end{aligned}
\label{eq:appK3}
\end{equation}
so that
\begin{equation}
\Gamma_3[X_1,X_2,A_t]
=\frac12{\rm Tr}_{t,{\rm adj}}\log{\cal K}_3.
\label{eq:appDet}
\end{equation}
The exact reduced functional is nonlocal.  At low temperature its nonzero-winding sectors obey, within the gapped MHO expansion,
\begin{equation}
\Gamma_3
=
\Gamma_{\rm vac}
+{\cal O}\!\left(e^{-\beta\Omega_3}\right),
\label{eq:appLowT}
\end{equation}
while the zero-winding determinant admits a local expansion around the MHO saddle inherited from the large-$d$ dynamics~\cite{YdriLatentTransport}.  For static $N=2$ configurations the surviving vacuum term can also be written exactly.  With
$Q={\rm Tr}(X_1^2+X_2^2)$ and $Y=-{\rm Tr}[X_1,X_2]^2$, one finds
\begin{equation}
\begin{aligned}
\frac{\Gamma_{\rm vac}^{\rm static}}{\beta}
=\frac12\Big[&\sqrt{m+2Q}
+\sqrt{m+Q+\sqrt{Q^2-2Y}}\\
&+\sqrt{m+Q-\sqrt{Q^2-2Y}}\Big].
\end{aligned}
\label{eq:appStaticDet}
\end{equation}
Thus the eliminated coordinate leaves an exact gauge-invariant functional of the two BFSS$_3$ invariants $Q$ and $Y$ before the subsequent local/Gaussian approximations.

\medskip
\noindent
The remaining two-matrix theory is then organized around an optimized
frequency $\omega_2$, following Pavel's finite-$N$ Gaussian description
of $SU(2)$ Yang--Mills quantum mechanics~\cite{PavelYM}.  Denoting by
$g^2$ the quartic Yang--Mills coupling in the reduced Hamiltonian and by
$\Omega_3^2$ the inherited quadratic scale, with $\Omega_3$ determined
by the preceding $d=3$ gap equation $\Omega_3^3-m\Omega_3=3$, the
leading gap equation is
\begin{equation}
\omega_2^3-\Omega_3^2\omega_2-g^2=0,
\qquad
\epsilon_{\rm G}\equiv\frac{g^2}{\omega_2^3}
\simeq\frac{g^2}{\Omega_3^3}.
\label{eq:appGap}
\end{equation}
Hence the finite-$N$ Gaussianization becomes parametrically controlled
in the large-mass regime.  Even in the massless case, where
$\epsilon_{\rm G}=1$, the optimized Gaussian vacuum has overlap
amplitude $0.973312$ with the exact ground state, corresponding to a
squared overlap of $0.94734$, as shown below; see
also~\cite{YdriLatentTransport}.

\medskip
\noindent
Once the two-matrix MHO is reached, no Gaussian approximation enters the final singlet projection.  For $d$ adjoint bosonic oscillators,
\begin{equation}
Z^{\rm bos}_{N,d}(x)=
\int_{SU(N)}d\mu(g)\,
\frac{1}{\det_{\rm adj}(1-x\,{\rm Ad}_g)^d}.
\label{eq:appMWgeneral}
\end{equation}
Here $x=e^{-\beta s}$, and the integral implements the Gauss law exactly~\cite{OConnor2023,OConnor2024}.  At $N=2$,
\begin{equation}
Z^{\rm bos}_{2,1}(x)=\frac{1}{1-x^2},
\qquad
Z^{\rm bos}_{2,2}(x)=\frac{1}{(1-x^2)^3}.
\label{eq:appMWvalues}
\end{equation}
The Hamiltonian reason is equally transparent.  Two adjoint $SU(2)\simeq SO(3)$ oscillator vectors $a_{aA}^\dagger$ admit precisely three independent quadratic singlet creators,
\begin{equation}
G_{11}=a_{1A}^\dagger a_{1A}^\dagger,
\qquad
G_{22}=a_{2A}^\dagger a_{2A}^\dagger,
\qquad
G_{12}=a_{1A}^\dagger a_{2A}^\dagger.
\label{eq:appGram}
\end{equation}
With only two flavor labels, the possible cubic $\epsilon_{ABC}$ invariant vanishes, while pairs of $\epsilon$ tensors reduce to Kronecker contractions.  Thus the singlet creation algebra is the free polynomial ring $\mathbb C[G_{11},G_{22},G_{12}]$~\cite{WeylClassicalGroups,ProcesiInvariantTheory}, giving Eq.~\eqref{eq:MWfactor} and
\begin{equation}
{\cal H}^{\rm sing}_{\rm BFSS_3-MHO}
\simeq
\left({\cal H}^{\rm sing}_{\rm BFSS_2}\right)^{\otimes3}.
\label{eq:appHilbertCube}
\end{equation}
This factorization occurs only after the common $SU(2)$ Gauss projection.

\section{Exceptional BFSS$_2$ decoder}
\label{app:decoder}

\medskip
\noindent
Let $M=N^2-1$ be the number of adjoint oscillator components.  In oscillator variables the gauge-invariant $\mathfrak{su}(1,1)$ generators may be written
\begin{equation}
\begin{aligned}
K_0&=\frac12\sum_{a=1}^{M}\left(a^{a\dagger}a^a+\frac12\right),\\
K_+&=\frac12\sum_{a=1}^{M}a^{a\dagger}a^{a\dagger},
\qquad K_-=K_+^\dagger.
\end{aligned}
\label{eq:appGenerators}
\end{equation}
with
\begin{equation}
\begin{aligned}
[K_0,K_\pm]&=\pm K_\pm,
\qquad [K_-,K_+]=2K_0,\\
C&=-K_0^2+\frac12(K_+K_-+K_-K_+).
\end{aligned}
\label{eq:appAlgebra}
\end{equation}
They commute with the $SU(N)$ Gauss generators and therefore act within the physical singlet Hilbert space.  The Fock vacuum is a lowest-weight singlet,
\begin{equation}
K_-|0\rangle=0,
\qquad
K_0|0\rangle=\frac{M}{4}|0\rangle,
\label{eq:appVacuumWeight}
\end{equation}
which gives
\begin{equation}
k_0=\frac{N^2-1}{4},
\qquad
C=-k_0(k_0-1).
\label{eq:appCasimir}
\end{equation}
For $N=2$ there is no independent primitive single-matrix invariant beyond the quadratic descendant tower.  Indeed, for a traceless $2\times2$ matrix $\Phi$, Cayley--Hamilton gives
\begin{equation}
\Phi^2=\frac12{\rm Tr}(\Phi^2){\bf 1}_2,
\label{eq:appCHSU2}
\end{equation}
so every invariant polynomial is generated by ${\rm Tr}(\Phi^2)\propto K_+$ and lies in the vacuum descendant tower.  The canonical lowest-weight Fock basis makes the module
$D^+_{3/4}$ manifest~\cite{YdriBFSS2Latent}.  This is one global
unitary realization of the same stable $N=2$ oscillator and does not
exhaust the unitary realizations admitted by its exceptional
$(k,C,\Lambda)$ data.

\medskip
\noindent
For a lowest-weight module,
\begin{equation}
K_-|k,k\rangle=0,
\qquad
C|k,m\rangle=-k(k-1)|k,m\rangle,
\label{eq:appLowestWeight}
\end{equation}
and repeated application of $K_+$ generates the full $D_k^+$ tower.  The value $k=3/4$ is exceptional.  One unitary realization is the lowest-weight discrete series $D^+_{3/4}$; for the universal cover $\widetilde{SL(2,\mathbb R)}$ there is no integrality obstruction to this value, and its quantized orbit gives $\mathbf{H}^2_\theta=\mathbf{EAdS}^2_\theta$.  At the same time,
\begin{equation}
0<\frac34<1,
\qquad
C=\frac{3}{16}\in\left(0,\frac14\right),
\label{eq:appComplementaryRange}
\end{equation}
which is also the complementary-series range in the same Casimir convention~\cite{Bargmann1947}.  The distinction is visible directly in the spectra of the compact generator:
\begin{equation}
\begin{array}{rcl}
D^+_{3/4}&:&K_0=\frac34+n,\qquad n=0,1,2,\ldots,\\[1mm]
C^0_{3/4}&:&K_0=\varepsilon+n,\qquad n\in\mathbb Z,
\end{array}
\label{eq:appRepSpectra}
\end{equation}
where the complementary series has no lowest-weight state (the
offset $\varepsilon$ depends on the global realization).  Hence the
same stable $N=2$ oscillator with
$k=3/4$, $C=3/16$ and $\Lambda=-s^2$ admits both the
lowest-weight discrete-series realization
$\mathbf{H}^2_\theta=\mathbf{EAdS}^2_\theta$ and the inequivalent
complementary-series realization $\mathbf{dS}^2_\theta$.  The two
are distinguished by representation structure, global realization
and $\ast$--structure, not by any change of oscillator dynamics.
They therefore stand on equal representation-theoretic footing; an
additional dynamical criterion would be needed only to privilege one
over the other.

\section{Rayleigh--Ritz test of the Gaussian approximation}\label{app:numerics}

\medskip
\noindent
The Gaussian/MHO description can be tested directly in the $N=2$,
$d=2$ theory by diagonalizing the full gauge-reduced Hamiltonian in an
MHO basis of arbitrary frequency $\omega$,
\begin{equation}
H
=
H_\omega
+\frac12(\Omega^2-\omega^2)r^2
+\frac{g^2}{16}r^4(1-x).
\label{eq:EM_num_split}
\end{equation}
The reduced singlet problem carries the physical measure
$d\mu_{\rm phys}=r^5\,dr\,dx$, while the Rayleigh--Ritz basis is
truncated by $2n+4\ell\leq K$, with $K$ the variational cutoff controlling
the size of the truncated Hilbert space.  Increasing $K$ therefore
systematically approaches the exact ground state.

\medskip
\noindent
Here $\Omega$ denotes the quadratic frequency of the reduced
two-matrix Hamiltonian.  In the latent chain one has
$\Omega=\Omega_3$, while $\Omega=0$ provides the massless
Yang--Mills benchmark.  The basis frequency $\omega$ is arbitrary,
whereas the optimized Gaussian frequency $\omega_{\rm opt}$ is the
positive solution of
$\omega^3-\Omega^2\omega-g^2=0$.  The one-state Rayleigh--Ritz
approximation is precisely the optimized Gaussian vacuum.

\medskip
\noindent
The resulting comparison between the optimized Gaussian approximation
and the Rayleigh--Ritz ground state is summarized in
Table~\ref{tab:EM_num_summary}.

\begin{table}[t]
  \centering
  \caption{Gaussian versus Rayleigh--Ritz ground-state results.}
\label{tab:EM_num_summary}
\begin{tabular}{|c|c|c|}
\hline
&
$g=1,\ \Omega=0$
&
$g=1,\ \Omega=1$
\\
\hline
$\omega_{\rm opt}$
& $1$
& $1.3247179572$
\\
\hline
$E_0$
& $2.1107035523$
& $3.5156457999$
\\
\hline
$E_G$
& $2.25$
& $3.5467736535$
\\
\hline
$(E_G-E_0)/E_0$
& $6.60\%$
& $0.8854\%$
\\
\hline
$|\langle\Phi_{00}|S_0\rangle|^2$
& $0.9473368022$
& $0.9940272952$
\\
\hline
\end{tabular}
\end{table}

The Rayleigh--Ritz energies converge monotonically with $K$ and are
independent, to the quoted precision, of the arbitrary basis frequency:
at $K=64$ the massless result is unchanged for
$\omega=0.8,1.0,1.2$, while the massive result is unchanged for
$\omega=1.0,1.3247179572,1.6$.  Thus the optimized MHO vacuum already
carries about $94.7\%$ of the exact ground-state norm in the massless
theory and about $99.4\%$ in the massive theory, while the corresponding
energy error decreases from $6.60\%$ to $0.885\%$.  This provides an
independent numerical check that the MHO description becomes
substantially more accurate once the inherited quadratic mass is
present.


\begin{thebibliography}{99}

\bibitem{BFSS}
T.~Banks, W.~Fischler, S.~H.~Shenker and L.~Susskind,
``M theory as a matrix model: A conjecture,''
Phys. Rev. D \textbf{55}, 5112 (1997), arXiv:hep-th/9610043.

\bibitem{BMN}
D.~Berenstein, J.~Maldacena and H.~Nastase,
``Strings in flat space and pp waves from $\mathcal N=4$ super Yang--Mills,''
JHEP \textbf{04}, 013 (2002), arXiv:hep-th/0202021.

\bibitem{Myers}
R.~C.~Myers,
``Dielectric-branes,''
JHEP \textbf{12}, 022 (1999), arXiv:hep-th/9910053.

\bibitem{Hoppe}
J.~Hoppe,
``Quantum theory of a massless relativistic surface and a two-dimensional bound state problem,''
Ph.D. thesis, MIT (1982).

\bibitem{Madore}
J.~Madore,
``The fuzzy sphere,''
Class. Quant. Grav. \textbf{9}, 69 (1992).

\bibitem{AzumaFuzzy}
T.~Azuma, S.~Bal, K.~Nagao and J.~Nishimura,
``Nonperturbative studies of fuzzy spheres in a matrix model with the Chern--Simons term,''
JHEP \textbf{05}, 005 (2004), arXiv:hep-th/0401038.

\bibitem{GeometryTransitionPRL}
R.~Delgadillo-Blando, D.~O'Connor and B.~Ydri,
``Geometry in transition: A model of emergent geometry,''
Phys. Rev. Lett. \textbf{100}, 201601 (2008), arXiv:0712.3011 [hep-th].

\bibitem{MatrixModelsEmergent}
R.~Delgadillo-Blando, D.~O'Connor and B.~Ydri,
``Matrix geometries and matrix models,''
JHEP \textbf{05}, 049 (2009), arXiv:0806.0558 [hep-th].

\bibitem{SteinackerEmergentGeometry}
H.~Steinacker,
``Emergent geometry and gravity from matrix models: an introduction,''
Class. Quant. Grav. \textbf{27}, 133001 (2010), arXiv:1003.4134 [hep-th].

\bibitem{MandalMahatoMorita}
G.~Mandal, M.~Mahato and T.~Morita,
``Phases of one dimensional large $N$ gauge theory in a 1/$D$ expansion,''
JHEP \textbf{02}, 034 (2010), arXiv:0910.4526 [hep-th].

\bibitem{FilevOConnorBFSS}
V.~G.~Filev and D.~O'Connor,
``The BFSS model on the lattice,''
JHEP \textbf{05}, 167 (2016), arXiv:1506.01366 [hep-th].

\bibitem{YdriLargeD}
B.~Ydri,
``Two approaches to quantum gravity and M-(atrix) theory at large number
of dimensions,''
Int.\ J.\ Mod.\ Phys.\ A {\bf 36}, no.~31n32, 2150234 (2021),
arXiv:2007.04488 [hep-th].

\bibitem{YdriLargeDSaddle}
B.~Ydri,
``A Double--Scaling Large--$d$ Saddle of BFSS/BMN Matrix Quantum Mechanics,''
Int.\ J.\ Mod.\ Phys.\ A (2026),
arXiv:2606.17758 [hep-th].

\bibitem{PavelYM}
H.-P.~Pavel,
``$SU(2)$ Yang--Mills quantum mechanics of spatially constant fields,''
Phys. Lett. B \textbf{648}, 97--106 (2007),
arXiv:hep-th/0701283.

\bibitem{BottcherWenzel}
A.~B\"ottcher and D.~Wenzel,
``The Frobenius norm and the commutator,''
Linear Algebra Appl. \textbf{429}, 1864 (2008).

\bibitem{YdriLatentTransport}
B.~Ydri,
``Latent Geometry: Geometric Transmutation and the Emergence of Lorentzian Time in BFSS/BMN,''
companion manuscript, BFSS/BMN Matrix Quantum Mechanics X (2026).

\bibitem{OConnor2023}
D.~O'Connor and S.~Ramgoolam,
``Gauged permutation invariant matrix quantum mechanics: path integrals,''
JHEP \textbf{04}, 080 (2024), arXiv:2312.12397 [hep-th].

\bibitem{OConnor2024}
D.~O'Connor and S.~Ramgoolam,
``Permutation invariant matrix quantum thermodynamics and negative specific heat capacities in large $N$ systems,''
JHEP \textbf{12}, 161 (2024), arXiv:2405.13150 [hep-th].

\bibitem{YdriMolienWeyl}
B.~Ydri,
``Molien--Weyl Singlet Counting and BFSS$_2$--Factorization in Gaussian Matrix QM,''
arXiv:2605.04621 [hep-th].

\bibitem{WeylClassicalGroups}
H.~Weyl,
\emph{The Classical Groups: Their Invariants and Representations}
(Princeton University Press, Princeton, 1946).

\bibitem{ProcesiInvariantTheory}
C.~Procesi,
\emph{Lie Groups: An Approach through Invariants and Representations}
(Springer, New York, 2007).

\bibitem{Park2005}
J.-H.~Park,
``A study of a noncritical osp(1$|$2,$\mathbb R$) M-theory,''
Nucl. Phys. B \textbf{745}, 123 (2006), arXiv:hep-th/0510070.

\bibitem{Kim2006}
N.~Kim and J.-H.~Park,
``Massive super Yang--Mills quantum mechanics: Classification and the relation to supermembrane,''
Nucl. Phys. B \textbf{759}, 249 (2006), arXiv:hep-th/0607005.

\bibitem{Bargmann1947}
V.~Bargmann,
``Irreducible unitary representations of the Lorentz group,''
Ann. Math. \textbf{48}, 568 (1947).

\bibitem{HoLiLargeN}
P.-M.~Ho and M.~Li,
``Large $N$ expansion from fuzzy AdS$_2$,''
Nucl. Phys. B \textbf{590}, 198 (2000), arXiv:hep-th/0005268.

\bibitem{JurmanSteinacker}
D.~Jurman and H.~Steinacker,
``2D fuzzy anti-de Sitter space from matrix models,''
JHEP \textbf{01}, 100 (2014), arXiv:1309.1598 [hep-th].

\bibitem{PinzulStern}
A.~Pinzul and A.~Stern,
``Noncommutative AdS$_2$/CFT$_1$ duality: The case of massless scalar fields,''
Phys. Rev. D \textbf{96}, 066019 (2017), arXiv:1707.04816 [hep-th].

\bibitem{YdriBFSS2Latent}
B.~Ydri,
``BFSS2 Quantum Gravity and Latent Geometry,''
companion manuscript, BFSS/BMN Matrix Quantum Mechanics VIII (2026).

\bibitem{KimNishimuraTsuchiya}
S.-W.~Kim, J.~Nishimura and A.~Tsuchiya,
``Expanding (3+1)-dimensional universe from a Lorentzian matrix model for superstring theory in (9+1) dimensions,''
Phys. Rev. Lett. \textbf{108}, 011601 (2012), arXiv:1108.1540 [hep-th].


\bibitem{Ydri:2021AdS2I}
B.~Ydri,
``The AdS$_\theta^2$/CFT$_1$ correspondence and noncommutative geometry I:
A QM/NCG correspondence,''
Int.\ J.\ Mod.\ Phys.\ A \textbf{37}, no.~13, 2250077 (2022),
arXiv:2108.13982 [hep-th].

\bibitem{Ydri:2021AdS2II}
B.~Ydri,
``The AdS$_\theta^2$/CFT$_1$ correspondence and noncommutative geometry II:
Noncommutative quantum black holes,''
Int.\ J.\ Mod.\ Phys.\ A \textbf{37}, no.~13, 2250078 (2022),
arXiv:2109.00380 [hep-th].

\bibitem{Bouraiou:2021AdS2III}
L.~Bouraiou and B.~Ydri,
``The AdS$_\theta^2$/CFT$_1$ correspondence and noncommutative geometry III:
Phase structure of the noncommutative AdS$_\theta^2\times S_N^2$,''
Int.\ J.\ Mod.\ Phys.\ A \textbf{37}, no.~13, 2250079 (2022),
arXiv:2109.01010 [hep-th].

\end{thebibliography}
\end{document}